\documentclass[aip,jcp,reprint,amsmath,amssymb,superscriptaddress]{revtex4-2}
\usepackage[T1]{fontenc}
\usepackage[utf8]{inputenc}
\usepackage{color}
\usepackage{array}
\usepackage{booktabs}
\usepackage{float}
\usepackage{multirow}
\usepackage{amsmath}
\usepackage{amsthm}
\usepackage{amssymb}
\usepackage{graphicx}
\usepackage{comment}

\makeatletter

\usepackage{hyperref}
\usepackage{siunitx}

\makeatother

\usepackage{babel}
\begin{document}
\title{Stochastic Tensor Contraction for Efficient MP2}

\author{Jiace Sun}
\author{Garnet Kin-Lic Chan$^{*}$}
\email{gkc1000@gmail.com}
\affiliation{Marcus Center for Theoretical Chemistry, California Institute of Technology, Pasadena, CA 91125, USA}
\affiliation{Division of Chemistry and Chemical Engineering, California Institute of Technology, Pasadena, CA 91125, USA}

\begin{abstract}
Second-order M{\o}ller--Plesset perturbation theory (MP2) is one of the simplest correlated wave-function methods, but its conventional $O(N^5)$ cost limits its application to large systems. Stochastic tensor contraction (STC) has recently appeared as a general technique to evaluate high-order tensor contractions in quantum chemistry. Here, we apply STC to the $O(N^5)$ bottleneck step of Laplace-transformed DF-MP2, reducing the computational scaling to an $O(N^3)$ deterministic setup cost and an $O(N^2)$ stochastic cost at fixed absolute error in the total energy, after an $O(N^4)$ preparation of the input tensor. The estimator is unbiased and provides a way to specify the target stochastic error by estimating the number of required samples before the full calculation. We implement the algorithm using a hybrid deterministic--stochastic evaluation strategy, with grouped index sampling, to reduce the computational prefactor. Over a range of benchmark molecules containing up to $\sim 7000$ basis functions, the complete STC-MP2 calculation becomes faster than DF-MP2 at about 600 basis functions, at a fixed absolute error of 0.2~kcal/mol, and is 7.7--10.0 times faster in the largest systems, where the STC treatment takes only about 5\% of the total time. For the isomerization energies of C$_{60}$ isomers, STC-MP2 reaches a given accuracy with 6--21 times lower CPU cost than a leading local MP2 method. Our results further substantiate the power of STC to serve as a general tensor-contraction engine for quantum chemistry.
\end{abstract}

\maketitle

\section{Introduction}
Electron correlation methods are among the central computational tools of molecular electronic-structure theory.\cite{szabo1996,helgaker2000,bartlett2007}
Second-order M{\o}ller--Plesset perturbation theory (MP2) is one of the simplest such methods and remains an important benchmark theory. It also serves as a component in other modern quantum chemistry methods,\cite{moller1934} such as
regularized MP2 variants,\cite{lee2018,shee2021,rettig2022,carterfenk2023} double-hybrid density functionals,\cite{grimme2006,martin2020} as well as being  used to correct higher-level local correlation approximations.\cite{nagy2018}

The canonical cost of MP2 scales as $O(N^5)$ with system size $N$. This leaves a substantial gap between the system sizes accessible to $O(N^3)-O(N^4)$ cost mean-field calculations, and those accessible to correlated calculations.
Devising efficient and accurate methods to evaluate the MP2 expressions is therefore an important task. 

A large body of work has sought to reduce the cost of MP2 by exploiting additional structure.
Density fitting (DF), or the resolution of the identity, factorizes the electron-repulsion integrals through an auxiliary basis. This greatly reduces the prefactor of MP2,\cite{feyereisen1993,weigend2002} but does not by itself remove the fifth-order scaling.
Local and pair-natural-orbital formulations exploit the spatial locality of dynamical correlation and can reach near-linear or linear scaling by restricting the orbital-pair spaces that are treated explicitly.\cite{pulay1983,saebo1993,neese2009,riplinger2013,pinski2015,nagysamu2016,wang2023sparsity,wang2023local,shi2024local}
Laplace-transform and atomic-orbital formulations transform the MP2 energy expression to reduce its formal scaling, down to linear with integral screening,\cite{almlof1991,haser1992,ayala1999,glasbrenner2020,schafer2017} and tensor hypercontraction and related low-rank factorizations reduce it to $O(N^4)$.\cite{hohenstein2012,lu2015,bangerter2021,lee2020isdf,haritan2025}
Stochastic formulations avoid the deterministic enumeration of all contributions and recover the exact result in expectation, with reported scalings from $O(N^3)$ down to linear.\cite{willow2012,neuhauser2013,takeshita2017}
However, methods with strongly reduced scaling typically come with large prefactors, which prevent their meaningful application except to very large systems.

In this work we address the fifth-order MP2 bottleneck using the recently introduced technique of stochastic tensor contraction (STC).\cite{sun2026stc}
STC evaluates a tensor contraction by importance sampling over its summation indices and, for arbitrary contractions, uses a loop-breaking strategy to construct an efficiently sampleable approximation to the ideal importance distribution.
After a one-time setup of probability tables, built from lower-cost contractions of the input tensors,
each stochastic sample has $O(1)$ cost, and 
for extensive quantities represented in a local basis, the analysis of Ref.~\onlinecite{sun2026stc} predicts a variance that scales as $O(N^2)$, so that a fixed absolute statistical error requires $O(N^2)$ samples.
In our previous work, this framework produced large speedups for expensive tensor contractions in CCSD(T), demonstrating that STC can substantially reduce the cost of high-order correlated calculations while retaining a directly controllable statistical error.\cite{sun2026stc}

With density fitting and a Laplace transform of the energy denominator, the $O(N^5)$ bottleneck step of MP2 is the evaluation of the direct and exchange terms from the dressed three-index tensor, while the preparation of this input tensor, by the DF transform and the Laplace dressing, costs $O(N^4)$.
We apply STC to the $O(N^5)$ bottleneck step.
It then has an $O(N^3)$ deterministic setup for the sampling tables and an $O(N^2)$ stochastic sampling cost at fixed absolute error.
We further demonstrate that the absolute cost is also small: in the largest cc-pVTZ and cc-pVQZ calculations we study, the STC treatment takes only about 5\% of the time of the complete STC-MP2 calculation, which is itself 7.7--10.0 times faster than DF-MP2 and 5--8 times faster than the DF-HF calculation on the same system.
At a fixed absolute error of 0.2~kcal/mol, STC-MP2 becomes faster than DF-MP2 at about 600 basis functions, much earlier than other reduced-scaling MP2 methods at comparable accuracy.
The remaining $O(N^4)$ tensor preparation cost can be reduced by utilizing locality, similar to local correlation methods, which we leave to future work.

Below we describe the theory and implementation, systematic scaling and error tests on quasi-one-dimensional linear alkane chains and quasi-two-dimensional hydrogen-terminated hexagonal boron nitride sheets, the isomerization energies of C$_{60}$ isomers, and a study of a set of large, chemically relevant, molecules. We also showcase the complementary strengths of the STC approach to local approximations, for which we compare to domain-based local pair natural orbital MP2~\cite{pinski2015,pinski2018} and to the recent local MP2 of Wang et al.\cite{wang2025precision} 

\section{Theory}
\label{sec:theory}
\subsection{Laplace-transformed DF-MP2 and the direct/exchange split}
\label{sec:laplace}

We begin from the spin-summed MP2 correlation energy in the density fitting (DF) approximation,
\begin{equation}
E_{\text{MP2}} = -\sum_{ijab} \frac{(ia|jb)\left[2(ia|jb)-(ib|ja)\right]}{\Delta_{ijab}},
\label{eq:mp2}
\end{equation}
where $i,j$ label occupied, and $a,b$ virtual, orbitals, $\Delta_{ijab} = \varepsilon_a + \varepsilon_b - \varepsilon_i - \varepsilon_j > 0$, and the two-electron integrals are factorized through an auxiliary basis $\{P\}$ of size $N_{\text{aux}} = O(N)$,
\begin{equation}
(ia|jb) = \sum_P R^P_{ia} R^P_{jb}, \quad
R^P_{ia} = \sum_{P'} (ia|P')\,(\boldsymbol{J}^{-1/2})_{P'P},
\label{eq:ri}
\end{equation}
with $J_{P'Q'} = (P'|Q')$ the Coulomb metric of the auxiliary basis.

The energy denominator couples all four orbital indices.
To obtain a more conveniently factorizable form, we use the numerical Laplace representation of the denominator,\cite{almlof1991,haser1992}
\begin{equation}
\frac{1}{\Delta} \approx \sum_{k=1}^{M} w_k e^{-\beta_k \Delta},
\label{eq:laplace}
\end{equation}
and refer to the resulting formulation as LT-MP2.
Its nodes $\beta_k$ and weights $w_k$ are obtained with the minimax algorithm of Refs.~\onlinecite{hackbusch} and \onlinecite{takatsuka2008}.
In these studies, $M = 6$--$8$ is found to be enough to converge the quadrature error for most systems.
Defining the exponential factors
\begin{equation}
\boldsymbol{U}^{\text{o}}(\beta) = e^{+\beta \boldsymbol{F}^{\text{o}}/2},
\quad
\boldsymbol{U}^{\text{v}}(\beta) = e^{-\beta \boldsymbol{F}^{\text{v}}/2},
\label{eq:expF}
\end{equation}
where $\boldsymbol{F}^{\text{o}}$ and $\boldsymbol{F}^{\text{v}}$ are the occupied and virtual blocks of the Fock matrix, we absorb them into the \emph{dressed tensors}
\begin{equation}
\tilde{R}^P_{ia}(\beta) = \sum_{i'a'}
U^{\text{o}}_{ii'}(\beta)\, R^P_{i'a'}\, U^{\text{v}}_{a'a}(\beta).
\label{eq:dressed}
\end{equation}
The correlation energy then becomes a sum of plain tensor contractions with no energy denominator,
\begin{equation}
E_{\text{MP2}} = -\sum_{k=1}^{M} w_k \Big[ 2J(\beta_k) - K(\beta_k) \Big],
\label{eq:split}
\end{equation}
with the direct (Coulomb-like) and exchange-like terms
\begin{align}
J(\beta) &= \sum_{ijab}\sum_{PQ}
\tilde{R}^P_{ia}\tilde{R}^P_{jb}\tilde{R}^Q_{ia}\tilde{R}^Q_{jb},
\label{eq:J}\\
K(\beta) &= \sum_{ijab}\sum_{PQ}
\tilde{R}^P_{ia}\tilde{R}^P_{jb}\tilde{R}^Q_{ib}\tilde{R}^Q_{ja}.
\label{eq:K}
\end{align}

The two terms differ only in the index pattern of the second pair of tensors, but this difference determines their cost when evaluated exactly.
In the direct term the pair $(ia)$ appears together in two tensors, and likewise $(jb)$, so the contraction factorizes through the auxiliary matrix $M_{PQ} = \sum_{ia} \tilde{R}^P_{ia}\tilde{R}^Q_{ia}$, $J = \sum_{PQ} M_{PQ}^2$, as Fig.~\ref{fig:diagram} shows, and requires $O(N^4)$ cost. The low scaling of the direct contraction is the motivation for spin-opposite-scaled (SOS)-MP2\cite{jung2004} which discards the exchange term completely.
The exchange term contains the `crossed' index pattern $(ia|jb)(ib|ja)$ (see Fig.~\ref{fig:diagram}). The contraction does not factorize in the same way, and costs $O(N^5)$ to evaluate exactly; it is the source of the fifth-order scaling of LT-MP2.
In LT-MP2, the evaluation of Eqs.~\ref{eq:J} and \ref{eq:K} is thus the $O(N^5)$ bottleneck step, after the $O(N^4)$ preparation of the input tensor by the DF transform of Eq.~\ref{eq:ri} and the Laplace dressing of Eq.~\ref{eq:dressed}. In this work we evaluate the bottleneck step stochastically, with the same algorithm for both terms, which we describe next.

\subsection{Stochastic evaluation of the direct and exchange terms}
\label{sec:stc}

\begin{figure*}
\centering
\includegraphics[width=0.75\linewidth]{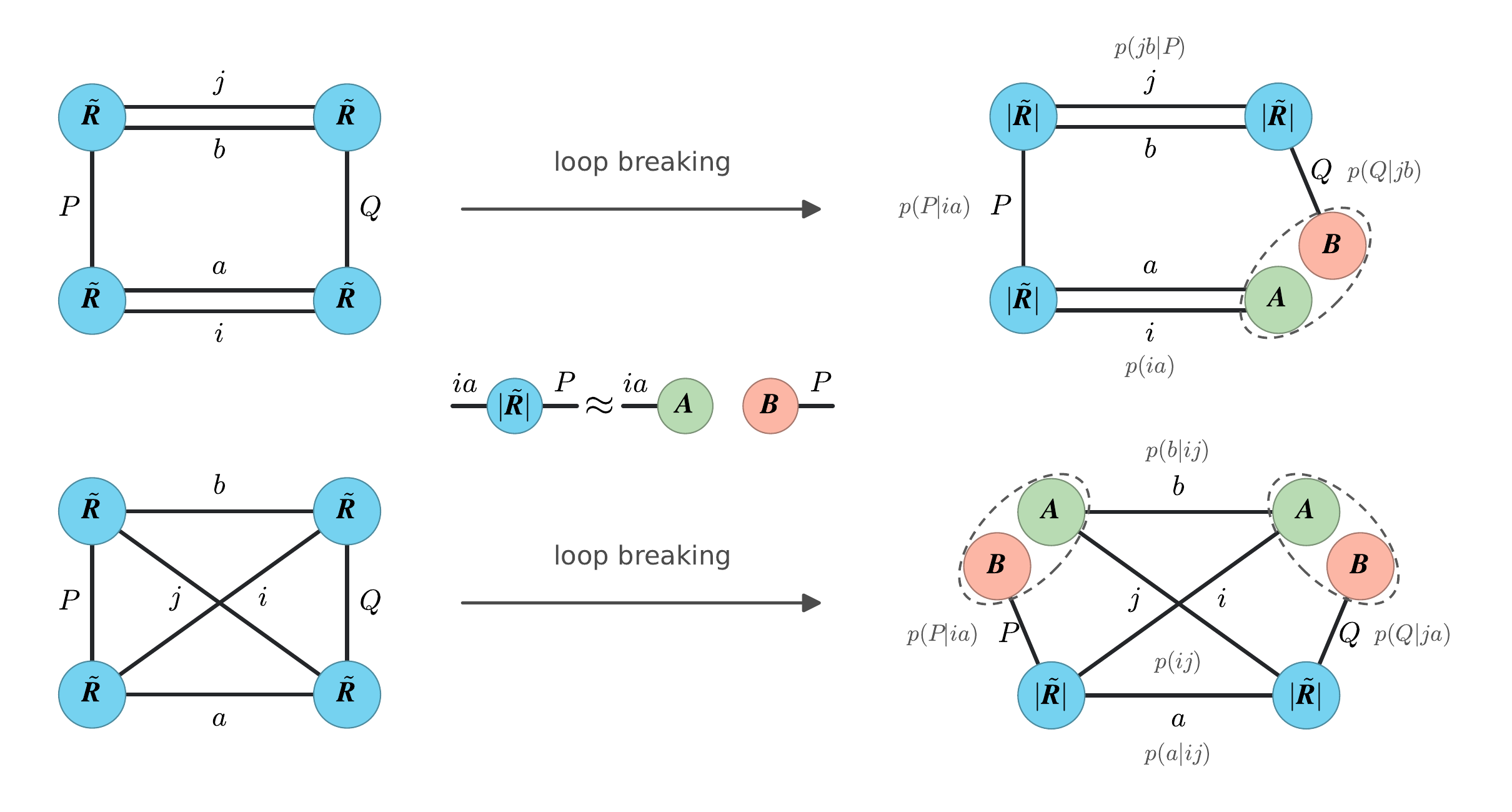}
\caption{Diagrams of the direct term $J$, Eq.~\ref{eq:J} (top left), and the exchange term $K$, Eq.~\ref{eq:K} (bottom left), in which each node is a dressed tensor $\tilde{\boldsymbol{R}}$ and each bond a summed index. In $J$ the pairs $ia$ and $jb$ are each shared by two tensors, so that the contraction factorizes through the auxiliary matrix $M_{PQ}$ and costs $O(N^4)$; in $K$ the crossed index pattern leads to an $O(N^5)$ contraction cost.
Loop breaking replaces tensors by the separable approximation $|\tilde{R}^P_{ia}| \approx A_{ia}B^P$ of Eq.~\ref{eq:guide_factors} (center), one tensor in $J$ and two in $K$ (dashed outlines), which gives the sampling guide probabilities of Eqs.~\ref{eq:guide_J} and \ref{eq:guide} (right).
Each index of the guides is labelled with the conditional from which it is drawn.}
\label{fig:diagram}
\end{figure*}

We evaluate Eqs.~\ref{eq:J} and \ref{eq:K} by stochastic tensor contraction (STC).\cite{sun2026stc}
The idea is to importance sample over the summation indices: for any sampling probability distribution $p$ over the composite index $I = (i,j,a,b,P,Q)$, the single-sample estimators
\begin{equation}
\begin{gathered}
\hat{J} = \frac{O^J_{I}}{p^J_{I}},
\quad
O^J_{I} = \tilde{R}^P_{ia}\tilde{R}^P_{jb}\tilde{R}^Q_{ia}\tilde{R}^Q_{jb},
\\
\hat{K} = \frac{O^K_{I}}{p^K_{I}},
\quad
O^K_{I} = \tilde{R}^P_{ia}\tilde{R}^P_{jb}\tilde{R}^Q_{ib}\tilde{R}^Q_{ja},
\end{gathered}
\quad
I \sim p,
\label{eq:estimator}
\end{equation}
are unbiased, $\mathbb{E}[\hat{J}] = J$ and $\mathbb{E}[\hat{K}] = K$, for any choice of $p$; the choice affects only the variance.
The optimal choice is $p_I \propto |O_I|$~\cite{sun2026stc}, but one requires an efficient way to sample from it. As discussed in Ref.~\onlinecite{sun2026stc}, this is simple if $O_I$ arises from a tree-like contraction, where $p_I$ can be obtained by sampling each of the tensors in the contraction and combining the conditionals (conditional probabilities).
However, both contractions are \emph{loopy}. In the direct term the four tensors form a ring (Fig.~\ref{fig:diagram}, top); in the exchange term every pair of the four tensors shares exactly one index, so its diagram is the complete graph of Fig.~\ref{fig:diagram}, bottom. In neither case does $|O_I|$ factorize into conditionals that can be sampled sequentially.
To address loopy contractions, Ref.~\onlinecite{sun2026stc} proposed a loop-breaking strategy, and we follow a similar procedure here.
We break the loops by replacing tensor factors with the separable approximation
\begin{equation}
\begin{gathered}
\big|\tilde{R}^P_{ia}\big| \approx A_{ia}\,B^P, \\
A_{ia} = \Big(\sum_P (\tilde{R}^P_{ia})^2\Big)^{1/2},
\quad
B^P = \Big(\sum_{ia} (\tilde{R}^P_{ia})^2\Big)^{1/2},
\end{gathered}
\label{eq:guide_factors}
\end{equation}
see Fig.~\ref{fig:diagram}.
In the direct term, replacing one factor, $\tilde{R}^Q_{ia}$, opens the ring into a chain. In the exchange term two factors, $\tilde{R}^P_{jb}$ and $\tilde{R}^Q_{ib}$, are replaced; this leaves a tree-like structure with a remaining loop (involving $i,j,a,b$), but we do not need to break this loop as the total probability can already be sampled efficiently.
Namely
\begin{align}
p^J_{ijabPQ} &= \frac{1}{Z^J}\,
\big|\tilde{R}^P_{ia}\big|\,\big|\tilde{R}^P_{jb}\big|\,\big|\tilde{R}^Q_{jb}\big|\;
A_{ia}\,B^Q,
\label{eq:guide_p_J}\\
p^K_{ijabPQ} &= \frac{1}{Z^K}\,
\big|\tilde{R}^P_{ia}\big|\,\big|\tilde{R}^Q_{ja}\big|\;
A_{ib}A_{jb}\,B^P B^Q,
\label{eq:guide_p}
\end{align}
with $Z^J$ and $Z^K$ being their normalizations, and these factorize into conditionals,
\begin{align}
p^J_{ijabPQ} &= p(ia)\,p(P|ia)\,p(jb|P)\,p(Q|jb),
\label{eq:guide_J}\\
p^K_{ijabPQ} &= p(ij)\,p(a|ij)\,p(b|ij)\,p(P|ia)\,p(Q|ja)
\label{eq:guide}
\end{align}
(see Fig.~\ref{fig:diagram} for how the tensors are grouped to define the conditionals).
Each conditional is tabulated once per quadrature point, at a cost proportional to the size of $\tilde{\boldsymbol{R}}$, after which indices are drawn in $O(1)$ time by the alias method.\cite{alias}
Thus, for each sample, evaluating Eq.~\ref{eq:estimator} requires only a small number of table lookups and tensor accesses.

The number of samples can then be determined from the requested accuracy. For each term we draw a small pilot batch of samples at every quadrature point, estimate the variance $\sigma^2$, and use $N_{\text{sample}} = \sigma^2/\epsilon^2$ samples for a target error $\epsilon$.
The requested variance of each term is distributed over the $M$ quadrature points a priori, in proportion to $w_k\,[\mathrm{Tr}\,\boldsymbol{U}^{\text{o}}(\beta_k)^2\,\mathrm{Tr}\,\boldsymbol{U}^{\text{v}}(\beta_k)^2]^{1.4}$, with the exponent chosen empirically, so that each quadrature point can be treated and the associated memory released before the next set of dressed tensors is formed.

When implementing the above, a few modifications are important for good performance.
First, drawing auxiliary indices one at a time means the per-sample work involves scattered memory accesses, which underutilizes modern computational hardware.
We therefore group the auxiliary functions by atom, sample the atom index $g$, and sum over each group exactly,
\begin{equation}
\sum_P \tilde{R}^P_{ia}\tilde{R}^P_{jb}
\;\longrightarrow\;
\sum_g \Big( \sum_{P\in g} \tilde{R}^P_{ia}\tilde{R}^P_{jb} \Big),
\label{eq:group}
\end{equation}
in both terms, with the sampling probabilities replaced by their group-summed counterparts. The per-sample computation is then a small dense contraction, and the variance is also reduced.
In practice a group contains $\sim$100 auxiliary functions for triple-zeta basis sets.

The second modification is an appropriate choice of basis. As discussed in Ref.~\cite{sun2026stc}, the STC variance is most favourable in a local representation. To use a local basis, the expressions in Eqs.~\ref{eq:expF}--\ref{eq:dressed} are evaluated in the canonical occupied and virtual spaces; and then the dressed tensor is transformed to the local representation used for sampling.
The occupied orbitals are localized with Pipek--Mezey.\cite{pipek1989}
For the virtual space, let $|\bar\chi_\mu\rangle$ denote the symmetrically orthogonalized atomic orbitals and $\hat P_{\mathrm v}$ the projector onto the virtual subspace.
We use the projected functions $|v_\mu\rangle = \hat P_{\mathrm v}|\bar\chi_\mu\rangle$, which remain atom-local and are overcomplete, but provide an exact resolution of the identity because
\begin{equation}
\sum_\mu |v_\mu\rangle\langle v_\mu| = \hat P_{\mathrm v}.
\label{eq:virtual_frame}
\end{equation}
Thus contractions over the virtual index are unchanged despite the redundant representation.
The auxiliary index is kept approximately local in the same spirit by using the symmetric inverse square root $\boldsymbol{J}^{-1/2}$ in Eq.~\ref{eq:ri} for orthogonalization.
We emphasize that STC is relatively insensitive to the precise method of localization: this is because localization is only used to reduce the variance, rather than to truncate contributions as in local correlation approximations. Indeed, we do not make any such local truncations in the current work.

With the estimator, the sampling distribution, and the grouping in place, we can now estimate the cost of the method.
As analyzed in Ref.~\onlinecite{sun2026stc}, the STC variance of an extensive quantity sampled in a local basis scales as $O(N^2)$, so that $N_{\text{sample}} = \sigma^2/\epsilon^2 = O(N^2)$ samples of $O(1)$ cost each suffices to reach a fixed absolute error $\epsilon$.
Therefore, for the bottleneck step, construction of the probability tables costs $O(N_{\text{occ}}N_{\text{vir}}N_{\text{aux}}) \sim O(N^3)$ and the stochastic sampling costs $O(N^2)$ at fixed absolute error, in place of the $O(N^5)$ of the deterministic evaluation.
It is worth noting that this sampling cost refers to a fixed absolute error. Many stochastic methods, such as sRI-MP2,\cite{takeshita2017} report their scaling for a fixed relative error, i.e.\ a fixed error per electron, which lowers the number of samples by an $O(N^2)$ factor; a fixed relative error, however, is typically not sufficient for molecular systems, where energy differences are required to chemical accuracy.
The input tensor preparation consists of the DF transform of Eq.~\ref{eq:ri}, which forms the three-index integrals, fits them, and transforms them to the occupied--virtual basis, and the Laplace dressing of Eq.~\ref{eq:dressed}, $O(N_{\text{occ}}N_{\text{vir}}N_{\text{aux}}(N_{\text{occ}}+N_{\text{vir}})) \sim O(N^4)$. Both are $O(N^4)$ dense matrix multiplications.
The memory usage is dominated by the dressed tensor, $N_{\text{occ}}N_{\text{vir}}N_{\text{aux}}$, as the quadrature points are processed one at a time.
The largest sampling tables are the auxiliary conditionals, which with the grouping of Eq.~\ref{eq:group} run over atoms rather than auxiliary functions and are therefore smaller than the dressed tensor by two orders of magnitude. In practice, we find the total memory requirement for systems studied in this work to be roughly $1.3$--$1.4$ times the size of a single $R_{ia}^P$ tensor.

\subsection{Hybrid deterministic--stochastic evaluation}
\label{sec:screen}

The efficiency of sampling versus deterministic summation in Eqs.~\ref{eq:J} and \ref{eq:K} depends on the distribution of sizes and number of terms in the sum.
The cost of deterministic evaluation (including local truncation) is proportional to the \emph{number} of terms in the summation, regardless of magnitude.
Stochastic evaluation costs $\sigma^2/\epsilon^2$, which is governed by the magnitude distribution of the sampled contributions rather than by how many there are; many tiny terms can be inexpensive to sample collectively, while a few dominant terms can strongly increase the variance.
The two techniques are therefore complementary. The most efficient arrangement is to treat the largest contributions deterministically and the many small ones stochastically.

We split the dressed tensor accordingly, using a threshold $\tau$ on the row norms of Eq.~\ref{eq:guide_factors}, scaled by the quadrature weight so that a single $\tau$ serves every quadrature point. Here $w$ denotes the quadrature weight at the current Laplace point.
We define the deterministic domain of occupied--virtual pairs and, for each occupied pair, the deterministic virtual domain
\begin{equation}
\begin{gathered}
D = \left\{ (ia) \,:\, w^{1/4} A_{ia} > \tau \right\},
\\
D_{ij} = \left\{ a \,:\, (ia) \in D \ \text{and} \ (ja) \in D \right\},
\end{gathered}
\label{eq:domain}
\end{equation}
and expand each of the two summations, over $(ia)$ and $(jb)$ in $J$ and over $a$ and $b$ in $K$, into a part inside and a part outside of the domain, to give the exact decompositions
\begin{align}
J &= \Bigg[
\sum_{\substack{(ia) \in D \\ (jb) \in D}}
+ \sum_{\substack{(ia) \notin D \\ (jb)}}
+ \sum_{\substack{(ia) \in D \\ (jb) \notin D}}
\Bigg]
\sum_{PQ} \tilde{R}^P_{ia}\tilde{R}^P_{jb}\tilde{R}^Q_{ia}\tilde{R}^Q_{jb},
\label{eq:hybrid_J}\\
K &= \sum_{ij}\Bigg[
\sum_{\substack{a \in D_{ij} \\ b \in D_{ij}}}
+ \sum_{\substack{a \notin D_{ij} \\ b}}
+ \sum_{\substack{a \in D_{ij} \\ b \notin D_{ij}}}
\Bigg]
\sum_{PQ} \tilde{R}^P_{ia}\tilde{R}^P_{jb}\tilde{R}^Q_{ib}\tilde{R}^Q_{ja}.
\label{eq:hybrid}
\end{align}
Each restricted summation is equivalent to an unrestricted one over a tensor whose entries outside the domain have been zeroed, so every term of Eqs.~\ref{eq:hybrid_J} and \ref{eq:hybrid} is a contraction of the same form as Eqs.~\ref{eq:J} and \ref{eq:K}, only with different input tensors; the estimators of Sec.~\ref{sec:stc} therefore apply to the sampled terms unchanged.
The first term is evaluated exactly, for $J$ as a single matrix product over the rows in $D$ and for $K$ as a small dense matrix product over $D_{ij}$, and the other two are sampled.
Note that Eqs.~\ref{eq:hybrid_J} and \ref{eq:hybrid} are identities for any $\tau$, because $\tau$ does not discard contributions or introduce bias, but only determines which part is computed deterministically versus stochastically.
It is therefore a pure efficiency parameter which does not need to be converged, unlike the thresholds in local correlation.

As $\tau$ grows the deterministic contribution shrinks and the variance of the sampled part rises, so the total cost passes through a minimum, which we analyze in Sec.~\ref{sec:threshold}. We use $\tau = 10^{-2}$ for the exchange term and $\tau = 5\times10^{-3}$ for the direct term in all production calculations in this work, except for the C$_{60}$ dimer in aug-cc-pVTZ (Sec.~\ref{sec:large}).

\section{Numerical results}
\label{sec:results}
\subsection{Computational details}
\label{sec:details}

The method of Sec.~\ref{sec:theory} was implemented in Python, using PySCF\cite{pyscf,pyscf2018} for the integrals and the mean-field reference, and PyTorch\cite{pytorch2019} with just-in-time (JIT) compilation and Numba\cite{numba2015} for the sampling kernels.
All calculations used a density-fitted restricted Hartree--Fock (DF-HF) reference and the frozen-core approximation, in the correlation-consistent cc-pVTZ basis\cite{dunning1989}, unless stated otherwise.
$M = 6$ minimax quadrature points were used in the Laplace transform, except for the C$_{60}$ isomers, for which we use $M = 8$. The resulting quadrature error is below $0.03\,\mathrm{m}E_h$ for the alkanes, the BN sheets, and the valinomycin and vancomycin macrocycles in cc-pVTZ, up to $0.2\,\mathrm{m}E_h$ for C3GC and $0.45\,\mathrm{m}E_h$ for the C$_{60}$ dimer in aug-cc-pVTZ, and negligible for the C$_{60}$ isomers with $M = 8$. We therefore assess the stochastic error against the exact LT-MP2 energy on the same quadrature, and report the total error against DF-MP2.
Unless stated otherwise, every STC calculation used requested standard errors of $0.3\,\mathrm{m}E_h\approx 0.19$~kcal/mol for the exchange term $K$ and $0.1\,\mathrm{m}E_h$ for the direct contribution $-2J$, i.e.\ $\sqrt{0.3^2+0.1^2} \approx 0.32\,\mathrm{m}E_h \approx 0.2$~kcal/mol in the correlation energy, and the screening thresholds $\tau = 10^{-2}$ for $K$ and $\tau = 5\times10^{-3}$ for $J$ (Sec.~\ref{sec:threshold}); no other parameter was adjusted between systems, except for the screening thresholds of the C$_{60}$ dimer in aug-cc-pVTZ (Sec.~\ref{sec:large}).

To illustrate the method, we use two families of systems with different effective dimensionality.
The first is a series of linear alkanes from C$_2$H$_6$ to C$_{100}$H$_{202}$, spanning 144 to 5828 basis functions and providing a quasi-one-dimensional setting.
The second is a series of hydrogen-terminated hexagonal boron nitride sheets, from BN 1$\times$1 to BN 8$\times$8, where BN $n\times m$ denotes an $n\times m$ array of the two-atom unit cell containing one boron and one nitrogen atom.
These span 264 to 5276 basis functions and provide a quasi-two-dimensional setting, which is a less local case.
Locality also depends on the HOMO--LUMO gap; the Hartree--Fock gap of the BN sheets (12.7--15.0~eV) is close to that of the alkanes (14.5--17.1~eV).
In addition, we study the isomerization energies of the C$_{60}$ isomers, with Hartree--Fock gaps down to 3.9~eV, in Sec.~\ref{sec:c60iso}, and a set of larger and chemically more realistic molecules with up to 6990 basis functions, including the C$_{60}$ dimer in the augmented aug-cc-pVTZ basis, in Sec.~\ref{sec:large}.

All the reported timings (including for comparison methods and packages) are wall times on eight CPU cores of a single Intel Xeon Platinum 8352Y CPU (2.20~GHz).
The STC-MP2 timings comprise the DF transform, the Laplace dressing, and the stochastic evaluation of $J$ and $K$.
Reference DF-HF, DF-MP2 calculations and timings were obtained with the standard PySCF implementation using the same auxiliary basis and frozen-core approximation.
DLPNO-MP2\cite{pinski2015,pinski2018} calculations were performed with ORCA\cite{orca,orca2012} using the TightPNO setting (TightPNO provides the more appropriate accuracy comparison than NormalPNO at these system sizes).
Unlike the present implementation of STC-MP2, DLPNO-MP2 operates with memory that grows linearly with system size.
For the reported errors, STC is referenced to PySCF DF-MP2 (or to the exact LT-MP2 energy, where stated), while DLPNO-MP2 is referenced to ORCA RI-MP2.

\subsection{Unbiasedness and error control}
\label{sec:unbiased}

\begin{figure}
\centering
\includegraphics[width=0.8\linewidth]{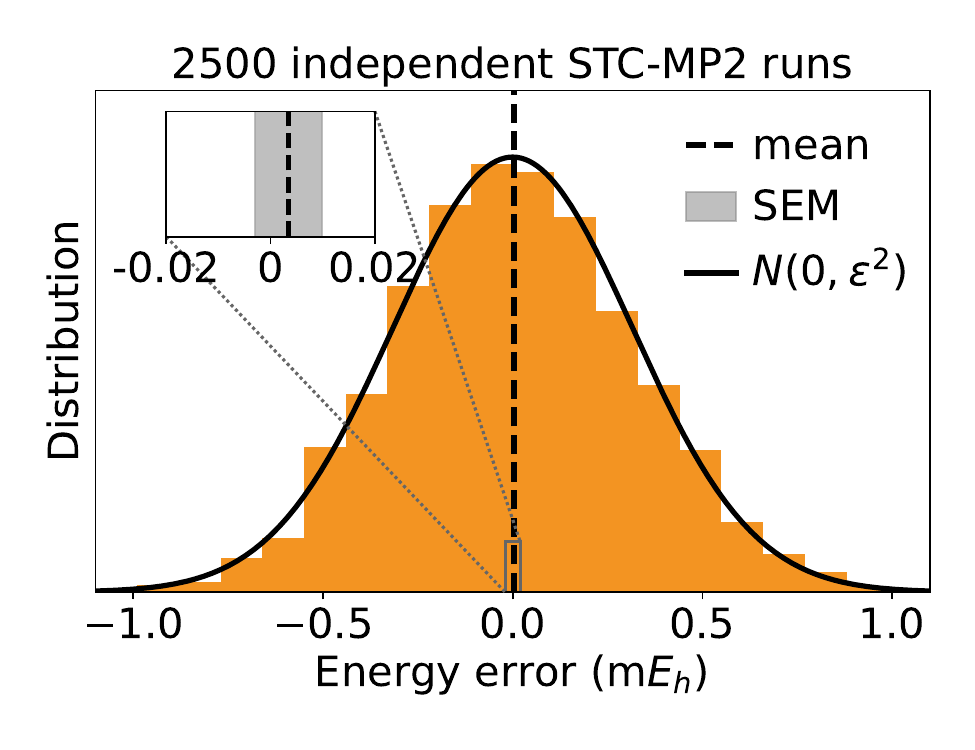}
\caption{Distribution of the STC-MP2 error over 2500 independent runs on BN 2$\times$2/cc-pVTZ, at requested standard errors of $0.3\,\mathrm{m}E_h$ for $K$ and $0.1\,\mathrm{m}E_h$ for $-2J$, i.e.\ $\epsilon = 0.32\,\mathrm{m}E_h$ in total. The reference is the exact LT-MP2 energy on the same quadrature, which differs from DF-MP2 by $0.002\,\mathrm{m}E_h$.
The solid curve is $N(0,\epsilon^2)$ with no fitted parameters. The dashed line is the sample mean.
The inset magnifies the sample mean and its standard error relative to zero.}
\label{fig:unbiasedness}
\end{figure}

The estimator of Sec.~\ref{sec:stc} is unbiased and we target a specified error through a sample count estimated from the variance, as described above.
We verify this error control numerically by repeating the same calculation 2500 times with independent random streams, on BN 2$\times$2/cc-pVTZ with the production settings of Sec.~\ref{sec:details}, i.e.\ a requested standard error of $\epsilon = \sqrt{0.3^2+0.1^2} \approx 0.32\,\mathrm{m}E_h$.
The resulting distribution of errors is shown in Fig.~\ref{fig:unbiasedness}.

The mean error is $+0.0034\,\mathrm{m}E_h$, which is 0.54 standard errors of the mean away from zero, so no statistically significant bias is detected at this sample size.
The measured standard deviation is $0.317\,\mathrm{m}E_h$, within 1\% of the prescribed $0.316\,\mathrm{m}E_h$, verifying that the variance-based sample count achieves the target statistical error.
As shown by the curve in Fig.~\ref{fig:unbiasedness}, which is $N(0,\epsilon^2)$ with $\epsilon=0.32$~m$E_h$, the errors closely follow the normal distribution, thus 
the error bar has the usual statistical interpretation,  and it can be reduced predictably as $1/\sqrt{N_{\text{sample}}}$ at a known cost.

\subsection{Deterministic–stochastic splitting}
\label{sec:threshold}

\begin{figure*}
\centering
\includegraphics[width=\linewidth]{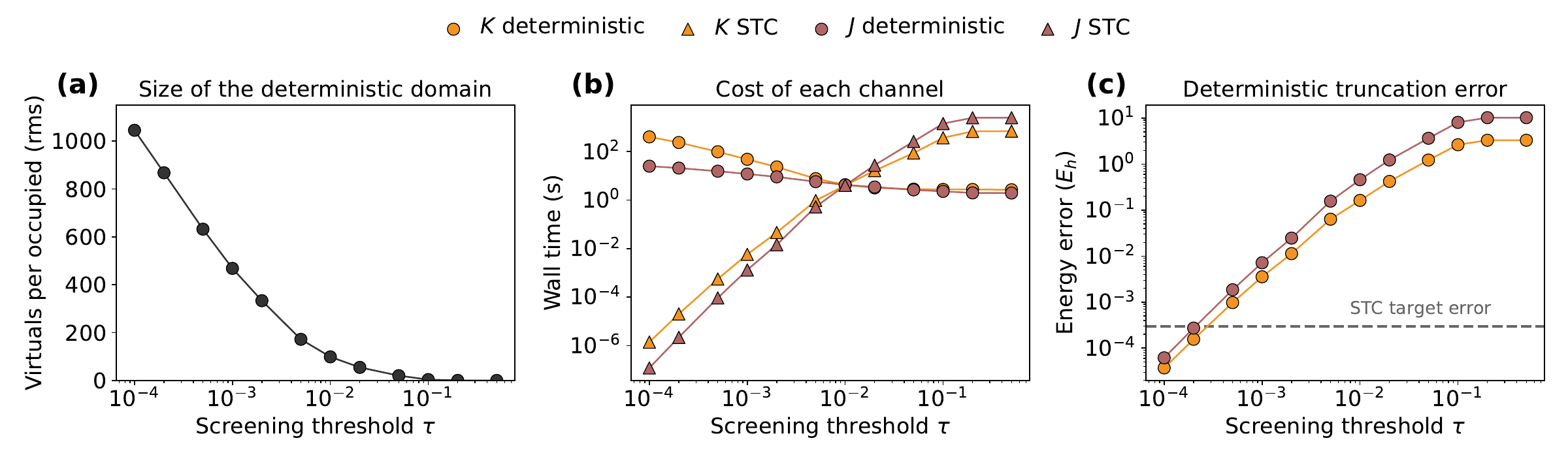}
\caption{Effect of the screening threshold $\tau$ on cost/accuracy of the deterministic and stochastic summations in the BN 4$\times$4 sheet with the cc-pVTZ basis, at a fixed requested accuracy of $0.3\,\mathrm{m}E_h$ for both $K$ (orange) and $-2J$ (red). The deterministic domain is shared by the two terms.
(a) Size of the domain for deterministic evaluation, as the root-mean-square number of projected virtual functions retained per occupied orbital.
(b) Wall time for the deterministic and the stochastic summations in Eqs.~\ref{eq:hybrid_J} and \ref{eq:hybrid}, on eight CPU cores.
(c) The deterministic truncation error, the error that would remain if the sampling contribution were neglected. The dashed line is the STC target error.}
\label{fig:threshold}
\end{figure*}

In Section~\ref{sec:screen} we argued that the deterministic and stochastic summations are efficient at capturing different aspects of the total contraction. 
We illustrate this in Figure~\ref{fig:threshold} for BN 4$\times$4/cc-pVTZ, where we change the threshold $\tau$ (that partitions the deterministic versus stochastic terms in the summation) over more than three orders of magnitude at a fixed requested accuracy of $0.3\,\mathrm{m}E_h$ for each term.
At every $\tau$ we measure the deterministic time and estimate the variance of the sampled part, from which the sampling time follows with the measured sampling rate.

Panel (a) shows that as $\tau$ is increased, the number of terms treated by the deterministic summation quickly drops. The root-mean-square number of projected virtual functions retained per occupied orbital falls from 1045 at $\tau = 10^{-4}$ to 98 at $\tau = 10^{-2}$ and to zero by $\tau = 2\times10^{-1}$.

Panel (b) shows the costs of the deterministic and stochastic parts of the algorithm separately.
As $\tau$ grows, the deterministic domain shrinks, and the deterministic time of $K$ falls from 399~s to a few seconds, while the variance of the remainder grows and the stochastic time rises to 688~s.
The total time of $K$ reaches a minimum of 8~s at $\tau = 10^{-2}$. This is 48 times faster than purely deterministic evaluation, and 84 times faster than purely stochastic evaluation.
The direct term behaves in the same way, with its minimum of 6~s also near $\tau = 10^{-2}$. Its purely deterministic evaluation is a single matrix product and costs only 25~s for this system, but grows as $O(N^4)$, whereas its purely stochastic evaluation is the most expensive, 2471~s.
The minima are shallow and lie near $\tau = 10^{-2}$ for both terms: between $\tau = 5\times10^{-3}$ and $10^{-2}$ the total time changes by less than 10\% for $K$ and 25\% for $J$. This suggests that the thresholds do not need to be carefully tuned per system. In practice we use the default $\tau = 10^{-2}$ for $K$ and $\tau = 5\times10^{-3}$ for $J$ for the remaining calculations, with requested standard errors of $0.3\,\mathrm{m}E_h$ for $K$ and $0.1\,\mathrm{m}E_h$ for $-2J$, i.e.\ $\sqrt{0.3^2+0.1^2} \approx 0.32\,\mathrm{m}E_h$ in total. 

Panel (c) shows the truncation error of a pure deterministic contraction as a function of $\tau$. 
At the thresholds used, this error is large ($162\,\mathrm{m}E_h$ for $K$ at $\tau = 10^{-2}$), and it only reaches the STC target error for $\tau \approx 2\times10^{-4}$, where from panel (b) the deterministic cost of $K$ is 234~s.
Thus, for this screening criterion and target accuracy, treating the discarded terms stochastically reduces the contraction time of $K$ by a factor of about 30.

\subsection{Sampling cost and its scaling}
\label{sec:samples}

\begin{figure}
\centering
\includegraphics[width=0.8\linewidth]{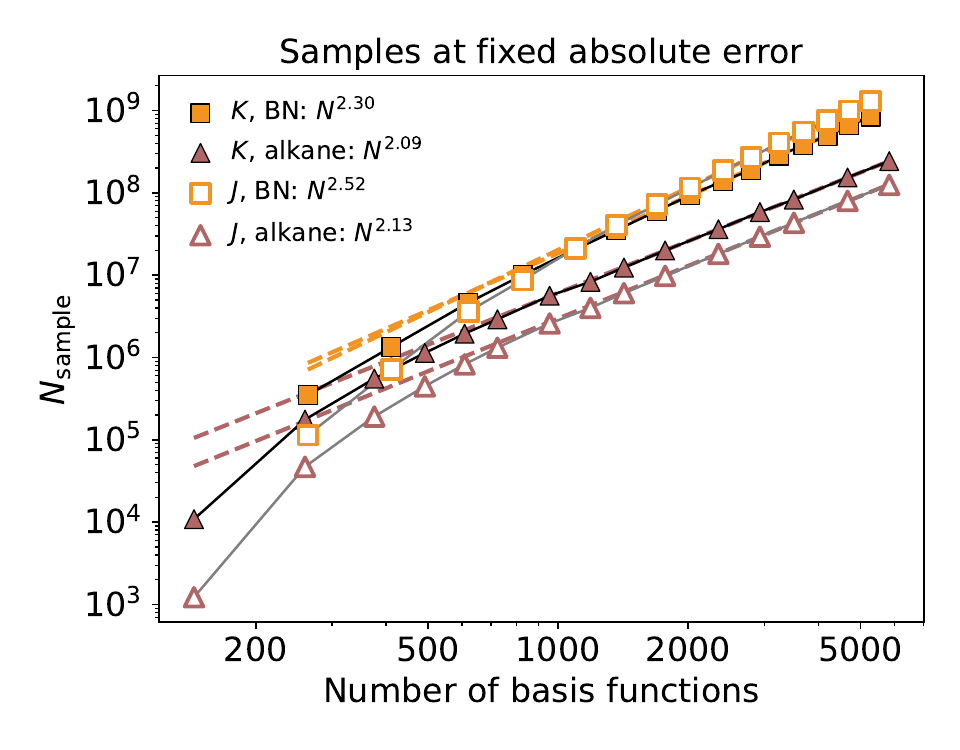}
\caption{Number of samples needed to reach absolute errors of $0.3\,\mathrm{m}E_h$ in the exchange term $K$ (filled symbols) and $0.1\,\mathrm{m}E_h$ in the direct contribution $-2J$ (open symbols), for 2D BN sheets and linear alkanes in cc-pVTZ.
Dashed lines are power-law fits over the largest systems, $N > 2000$, extrapolated across the full range.}
\label{fig:samples}
\end{figure}

In Sec.~\ref{sec:stc}, we claimed that a fixed absolute accuracy requires $O(N^2)$ samples of $O(1)$ cost each, in place of the $O(N_{\text{occ}}N_{\text{vir}}N_{\text{aux}}^2)$ and $O(N_{\text{occ}}^2 N_{\text{vir}}^2 N_{\text{aux}})$ operations of the exact direct and exchange contractions.
We demonstrate this in Figure~\ref{fig:samples}, recording the number of samples each calculation chooses in order to meet the requested accuracy of $0.3\,\mathrm{m}E_h$ for $K$ and $0.1\,\mathrm{m}E_h$ for $-2J$.

We observe that the sample count grows as $N^{2.09}$ for $K$ and $N^{2.13}$ for $J$ in the linear alkanes, in close agreement with the predicted $O(N^2)$, and as $N^{2.30}$ and $N^{2.52}$ for the 2D BN sheets.
The BN fits are somewhat steeper, which we attribute to the limited accessible size range.
Note that this scaling is for fixed \emph{absolute} error. Since the correlation energy is extensive, a fixed relative error permits an absolute error proportional to $N$; combined with the $O(N^2)$ variance, the sample count for a fixed relative error is therefore asymptotically $O(1)$.
The cost per sample is essentially constant, and about the same for the two terms, so up to a constant factor
Fig.~\ref{fig:samples} also gives the wall time of the stochastic contractions.

\subsection{Comparison with DF-MP2 and local MP2}
\label{sec:scaling}

\begin{figure*}
\centering
\includegraphics[width=\linewidth]{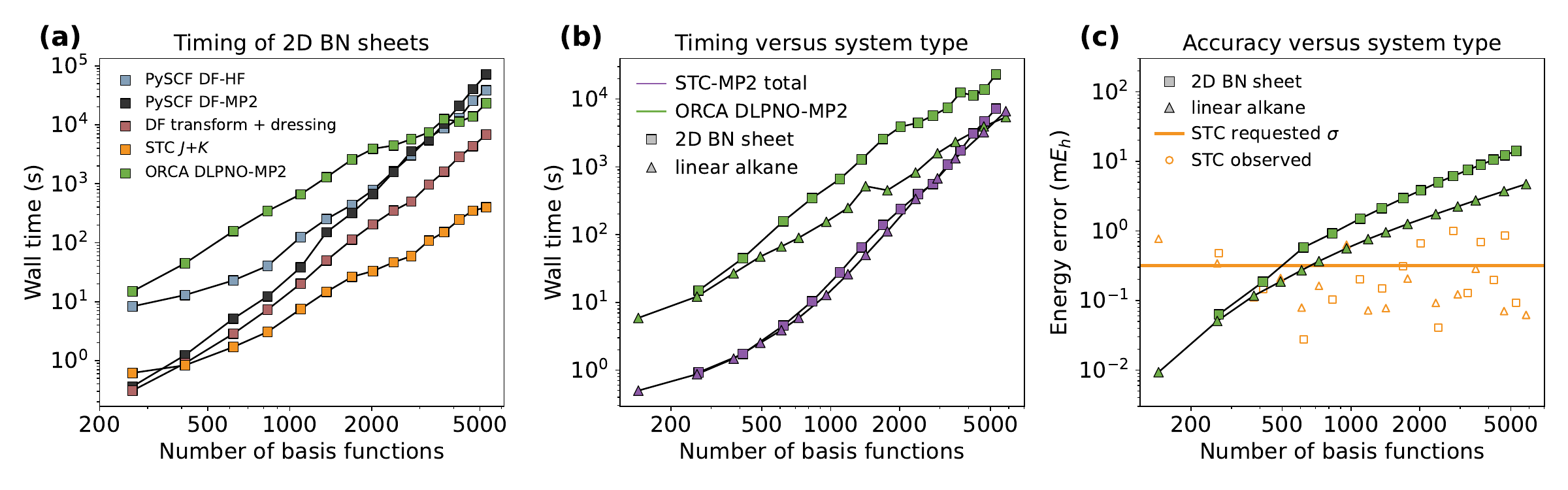}
\caption{Comparison of computational cost and accuracy for the present approach, standard DF-MP2, and DLPNO-MP2, in the cc-pVTZ basis and at a requested STC standard error of $0.32\,\mathrm{m}E_h$; all timings are on eight CPU cores.
(a) Wall times of the complete DF-HF and DF-MP2 calculations, of DLPNO-MP2, and of the two parts of STC-MP2 for the 2D BN sheets: the preparation of the dressed tensors (DF transform and Laplace dressing) and the stochastic evaluation of $J$ and $K$.
(b) Wall time of the complete STC-MP2 calculation and of DLPNO-MP2 at the TightPNO setting, for both system families.
(c) Error characterization of STC-MP2 and DLPNO-MP2. For STC the line is the requested one-standard-deviation statistical uncertainty, which is met by construction; the open symbols are the observed errors with respect to DF-MP2, shown to illustrate that they are consistent with it. For DLPNO-MP2 the values are the observed deterministic deviations from the DF-MP2 energy of the same package.}
\label{fig:scaling}
\end{figure*}

Figure~\ref{fig:scaling} compares STC-MP2 with standard DF-MP2 and DLPNO-MP2 in both computational cost and accuracy. Panel (a) shows the timing breakdown for the BN sheets, while panels (b) and (c) compare the timing of the complete calculations and the errors, respectively, of STC-MP2 and DLPNO-MP2 across the alkane and BN series.

Panel (a) shows the BN sheets only, since the costs of the deterministic methods and steps (DF-HF, DF-MP2, the DF transform, and the Laplace dressing) are determined by the system size and hardly depend on the system type; the alkanes give nearly the same curves. It shows a clear separation in the growth of the different costs with system size. Standard DF-MP2 grows most rapidly, reflecting its $O(N^5)$ scaling. DF-HF and the preparation of the dressed tensors grow more slowly, consistent with their $O(N^4)$ costs, while the stochastic evaluation of $J$ and $K$ grows more slowly still, and has the smallest cost for every system beyond $\sim 500$ basis functions.
For the largest BN system (8$\times$8) with 5276 basis functions, DF-MP2 takes 71752~s and the complete DF-HF calculation 38463~s. The complete STC-MP2 calculation takes 7204~s, of which the DF transform takes 2733~s, the Laplace dressing 4074~s, and the stochastic evaluation of $J$ and $K$ only 398~s (242~s and 156~s).
STC-MP2 is thus 10.0 times faster than DF-MP2, 3.2 times faster than DLPNO-MP2 with the TightPNO setting (23161~s), and 5.3 times faster than the DF-HF calculation on the same system.
Of the complete STC-MP2 time, the STC treatment of the $O(N^5)$ bottleneck step takes only 5.5\%; the remaining 94.5\% is the input tensor preparation, which consists of dense $O(N^4)$ matrix multiplications.
The complete STC-MP2 calculation becomes faster than DF-MP2 at about 600 basis functions, between C$_{8}$H$_{18}$ and C$_{10}$H$_{22}$ for the alkanes and between BN 1$\times$2 and BN 2$\times$2 for the sheets, i.e.\ at about 60 correlated electrons. Below this size its cost is dominated by the fixed overheads of the preparation step. This crossover is much earlier than for other reduced-scaling MP2 methods at comparable accuracy: the local MP2 of Wang et al.\ matches the timing of canonical RI-MP2 for systems of 2800--4000 basis functions in cc-pVTZ, at an RMSE of 0.16~kJ/mol in binding energies,\cite{wang2025precision} and DLPNO-MP2 with the TightPNO setting, at a much larger error, achieves crossover at about 2300 basis functions for the alkanes and above 3700 for the BN sheets in our calculations.
STC requires only the dressed tensors as input and is independent of how they are prepared, so the cost of this step can be reduced by the same local truncations as in local correlation methods, which we do not use here.

Panel (b) compares the complete STC-MP2 and DLPNO-MP2 calculations across the quasi-one-dimensional alkane chains and quasi-two-dimensional BN sheets. The STC-MP2 timings of the two system families nearly coincide at the same number of basis functions, despite their different dimensionality and locality, whereas DLPNO-MP2 is considerably more expensive for the 2D system, since the number of virtual orbitals in a correlation domain is exponential in the dimension.
DLPNO-MP2 grows more slowly with system size, so the comparison depends on the system: for C$_{100}$H$_{202}$, DLPNO-MP2 (5407~s) is faster than STC-MP2 (6629~s), whose cost is dominated by the dense preparation step, whereas for BN 8$\times$8 STC-MP2 is three times faster.

Panel (c) shows the corresponding accuracy comparison. The TightPNO error is $4.7\,\mathrm{m}E_h$ for the C$_{100}$ alkane and $14\,\mathrm{m}E_h$ for BN 8$\times$8, while the requested STC standard error is $0.32\,\mathrm{m}E_h$ for every system; the observed STC errors scatter around it as expected for a normal distribution, within about $3\sigma$. For the BN 8$\times$8 system the TightPNO error is therefore $\sim 45\times$ larger than the requested STC error. Errors in energy differences can be smaller than these total errors; we compare energy differences directly in Sec.~\ref{sec:c60iso}.

\subsection{Relative energies: C$_{60}$ isomerization}
\label{sec:c60iso}

\begin{figure}
\centering
\includegraphics[width=0.9\linewidth]{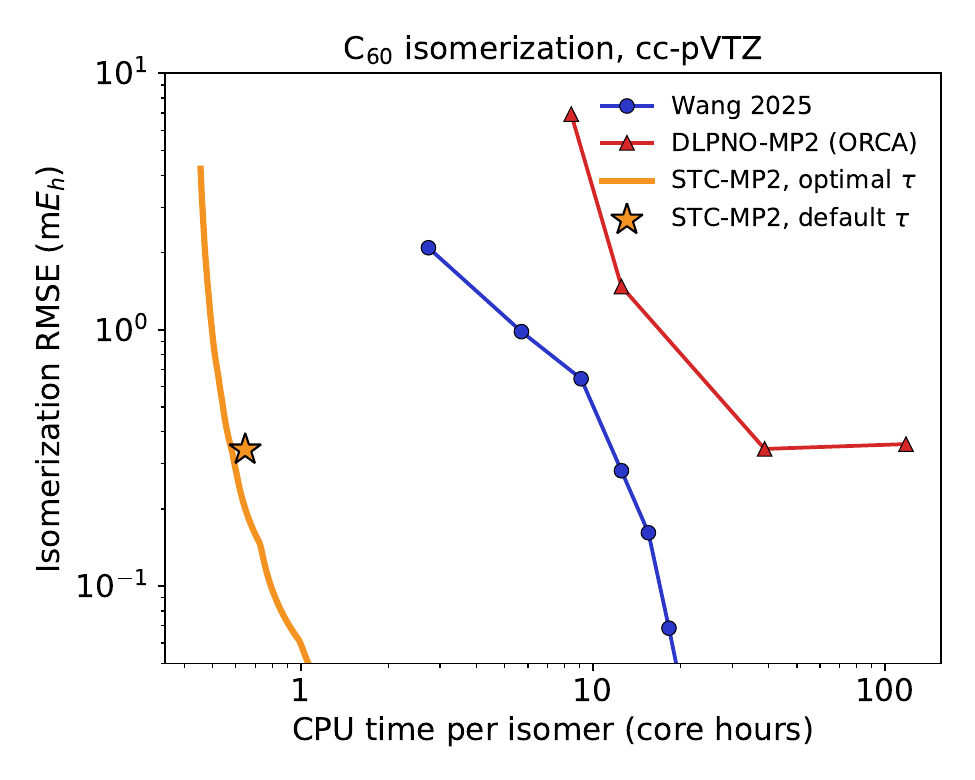}
\caption{CPU time per isomer against the RMSE of the nine C60ISO isomerization energies (cc-pVTZ), down to 0.05~m$E_h$. STC-MP2 ($M = 8$): cost--accuracy curve with the screening thresholds and the split of the error between $J$ and $K$ re-optimized at every target error (``optimal $\tau$'', line), obtained from the measured deterministic times, per-sample variances and sampling rates; and the production calculation with the default settings of Sec.~\ref{sec:details} (``default $\tau$'', star), with its error relative to DF-MP2. Local MP2 of Wang et al.\ and DLPNO-MP2: data adapted from Fig.~11 of Ref.~\onlinecite{wang2025precision}; copyright 2025 American Chemical Society. CPU time is the number of cores times the wall time: 32 threads on two AMD EPYC 7763 (2.45~GHz) for Ref.~\onlinecite{wang2025precision}, and eight cores of an Intel Xeon Platinum 8352Y (2.20~GHz) here.}
\label{fig:c60iso}
\end{figure}

The quantities of chemical interest are energy differences. For an unbiased stochastic method, the errors of two independent calculations add rather than cancel, so that the expected error of an energy difference is $\sqrt{2}\sigma$ for a per-calculation standard error of $\sigma$.
To test this on chemically relevant energy differences, we compute the C60ISO set of GMTKN55,\cite{goerigk2017gmtkn55} ten C$_{60}$ isomers (1800 basis functions in cc-pVTZ) and nine isomerization energies relative to the most stable isomer. This set and basis were used by Wang et al.\ to benchmark their local MP2 method against PNO methods.\cite{wang2025precision}

To obtain the cost as a function of the target error, we use $M = 8$ quadrature points, for which the quadrature error of the isomerization energies is negligible (RMSE 0.002~m$E_h$), so that the RMSE is the stochastic $\sqrt{2}\sigma$. For each isomer we measure the deterministic times and the per-sample variances of $J$ and $K$ at a range of screening thresholds, and combine them with the sampling rates measured in the production calculations; the thresholds and the split of the error between $J$ and $K$ are then chosen to minimize the mean time over the isomers. The resulting curve is shown in Fig.~\ref{fig:c60iso}. The CPU time per isomer is 0.47~CPU hours at an RMSE of 2~m$E_h$, 0.61 at 0.26~m$E_h$, 0.88 at 0.076~m$E_h$, and 1.06 at 0.05~m$E_h$. At large target errors the cost approaches that of the deterministic steps (DF transform and Laplace dressing, 0.41~CPU hours).
The production calculation with the settings of Sec.~\ref{sec:details} gives an RMSE of the nine isomerization energies of 0.34~m$E_h$ (0.89~kJ/mol) with respect to DF-MP2, at 0.65~CPU hours per isomer, consistent with the expected $\sqrt{2}\sigma = 0.45$~m$E_h$. Each point of the optimal curve requires its own thresholds and error split, whereas the production calculation uses the default settings of all other systems; it therefore takes about 17\% longer than the optimal curve at the same requested error.

Fig.~\ref{fig:c60iso} also shows the cost--accuracy curves reported by Wang et al.\ for their local MP2 and for DLPNO-MP2.\cite{wang2025precision} Since the calculations were run on different hardware with different numbers of cores, we compare total CPU times. At an RMSE of 0.26~m$E_h$, STC-MP2 needs 0.61~CPU hours per isomer, against about 12.5~CPU hours for 0.28~m$E_h$ with the local MP2 of Wang et al. Over the range from 2~m$E_h$ down to 0.05~m$E_h$, STC-MP2 needs 6--21 times less CPU time at the same RMSE.

\subsection{Benchmark on large realistic molecules}
\label{sec:large}

\begin{figure*}
\centering
\includegraphics[width=0.8\linewidth]{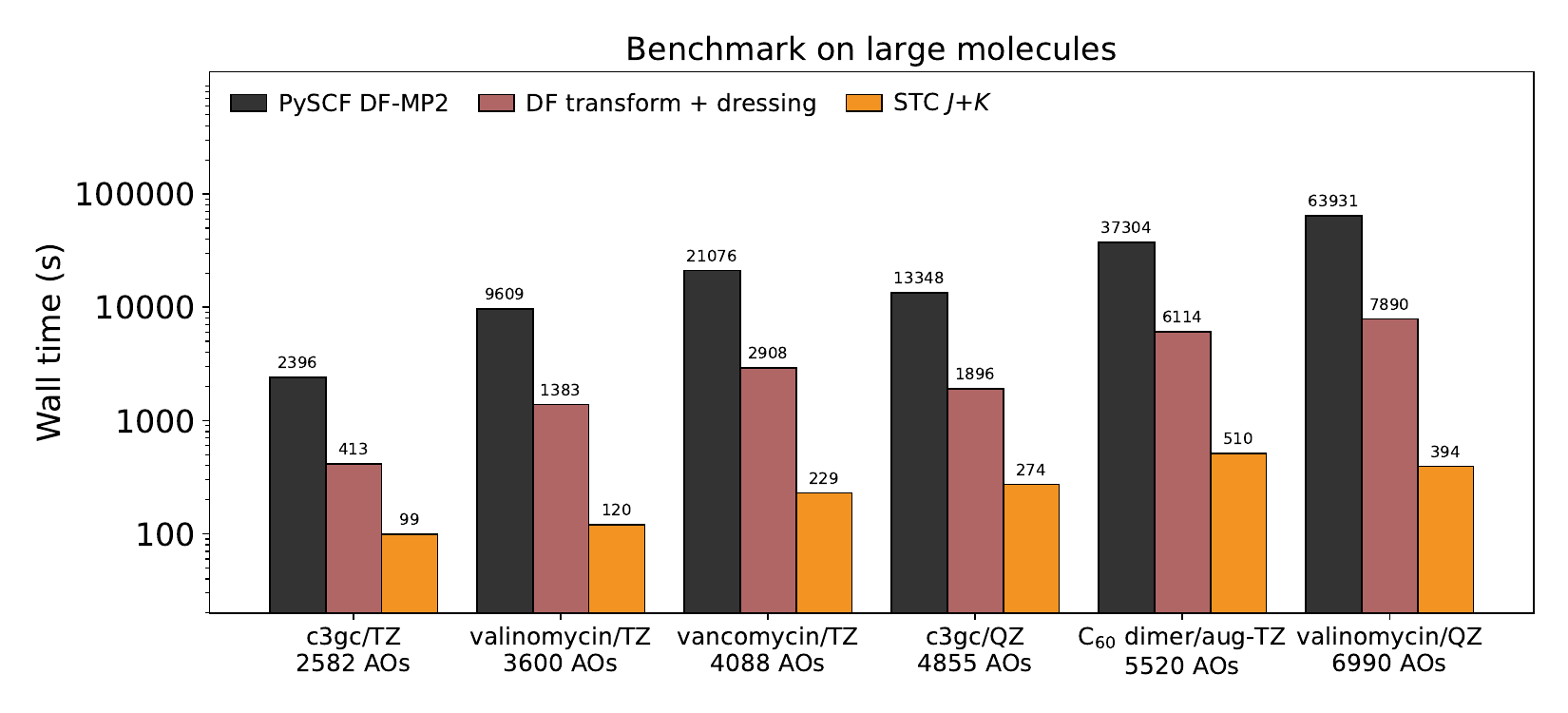}
\caption{Wall time on eight CPU cores for six large molecules, ordered by the number of atomic orbitals, at a requested STC standard error of 0.2~kcal/mol.
TZ, QZ, and aug-TZ denote the cc-pVTZ, cc-pVQZ, and aug-cc-pVTZ basis sets.\cite{dunning1989}
The preparation of the dressed tensors (DF transform and Laplace dressing) and the stochastic evaluation of $J$ and $K$ together give the complete STC-MP2 calculation.
Note the logarithmic scale for wall time.}
\label{fig:large}
\end{figure*}

We now consider six calculations on more `chemically realistic' molecules: the C3GC complex of the L7 set,\cite{sedlak2013} 101 atoms, and the macrocycles valinomycin, 168 atoms, and vancomycin, 176 atoms, in the cc-pVTZ basis, and, for the two smaller ones, the cc-pVQZ basis\cite{dunning1989} as well; and the C$_{60}$ dimer, 120 atoms, in the aug-cc-pVTZ basis, to test a basis with diffuse functions.
These contain 2582 to 6990 basis functions. For the largest calculation, we need 181~GB to store a single dressed tensor.

In Figure~\ref{fig:large}, we show the timings of a complete DF-MP2 calculation, of the preparation of the dressed tensors, and of the stochastic evaluation of $J$ and $K$. Table~\ref{tab:breakdown} breaks down the timings further for C3GC in cc-pVTZ and valinomycin in cc-pVQZ.

\begin{table*}
\caption{Timings (s, eight CPU cores) of STC-MP2 for C3GC/cc-pVTZ and valinomycin/cc-pVQZ at a requested standard error of 0.2~kcal/mol. ``DF'' is the DF transform and ``dressing'' the Laplace dressing. For each of $J$ and $K$, ``det.'' is the exact evaluation of the deterministic domain, ``tables'' the construction of the deterministic domain and the sampling tables, and ``sampling'' the drawing and evaluation of the samples.}
\label{tab:breakdown}
\setlength{\tabcolsep}{7pt}
\begin{tabular}{l@{\hspace{1.5em}}cc@{\hspace{1.5em}}cccc@{\hspace{1.5em}}cccc@{\hspace{1.5em}}c}
\toprule
 & \multicolumn{2}{c}{preparation} & \multicolumn{4}{c}{STC-$J$} & \multicolumn{4}{c}{STC-$K$} & \\
\cmidrule(r{1em}){2-3} \cmidrule(r{1em}){4-7} \cmidrule(r{1em}){8-11}
system & DF & dressing & det. & tables & sampling & total & det. & tables & sampling & total & STC-MP2 \\
\midrule
C3GC/TZ & 203 & 211 & 16 & 6 & 29 & 51 & 5 & 12 & 31 & 48 & 512 \\
valinomycin/QZ & 3713 & 4177 & 146 & 58 & 44 & 248 & 14 & 99 & 33 & 146 & 8284 \\
\bottomrule
\end{tabular}
\end{table*}

Across the five cc-pVTZ and cc-pVQZ calculations, the stochastic evaluation takes 99~s to 394~s, i.e.\ a factor of 4.0, to be compared with a factor of 27 for standard DF-MP2 over the same molecules, reflecting the change in scaling. For the largest system, valinomycin in cc-pVQZ, it takes 5\% of the complete STC-MP2 time of 8284~s, which is 7.7 times faster than DF-MP2 (63931~s) and 8.2 times faster than the DF-HF calculation (67541~s).
From triple- to quadruple-zeta, the stochastic evaluation grows by a factor of 2.8 for C3GC and 3.3 for valinomycin, while the preparation cost grows by 4.6 and 5.7, following its $O(N^4)$ scaling.
For the C$_{60}$ dimer in aug-cc-pVTZ (5520 basis functions), the diffuse functions reduce the locality of the orbitals and increase the variance: the production thresholds are no longer optimal. A scan of the screening thresholds gives $\tau = 5\times10^{-3}$ for $K$ and $3\times10^{-3}$ for $J$, which reduce the time of the stochastic evaluation by about 38\%, to 510~s. This is more than for the larger valinomycin calculation in cc-pVQZ, but still only 8\% of the complete STC-MP2 time of 6624~s, which is 5.6 times faster than DF-MP2 (37304~s).
The observed errors with respect to DF-MP2 are $+0.60$, $-0.10$, $-0.81$, $-0.01$, $-0.23$, and $+0.57\,\mathrm{m}E_h$ for C3GC/TZ, valinomycin/TZ, vancomycin/TZ, C3GC/QZ, the C$_{60}$ dimer/aug-TZ, and valinomycin/QZ, within the expected range of Gaussian fluctuations.
In this formulation the bottleneck is therefore no longer the $O(N^5)$ step, but the $O(N^4)$ input tensor preparation.
At the same time, storing the complete dressed tensor requires $O(N^3)$ memory, which is already substantial at the largest sizes considered here.
These limitations suggest natural extensions of the present approach, which we discuss in Sec.~\ref{sec:conclusion}.

\section{Conclusion and outlook}
\label{sec:conclusion}
We have applied stochastic tensor contraction to the $O(N^5)$ bottleneck step of Laplace-transformed DF-MP2, the evaluation of its direct and exchange terms, with the same algorithm for both.
Using this technique, this step requires an $O(N^3)$ deterministic setup of the sampling tables and an $O(N^2)$ stochastic sampling cost for a fixed absolute error.
The numerical results show that, at an absolute accuracy of 0.2~kcal/mol, the STC treatment takes only about 5\% of the time of the complete STC-MP2 calculation in the largest systems, which is 7.7--10.0 times faster than DF-MP2 and 5--8 times faster than the DF-HF calculation on the same system, and STC-MP2 becomes faster than DF-MP2 already at about 600 basis functions.
The errors are unbiased and add in quadrature in energy differences; for the C$_{60}$ isomerization energies, STC-MP2 needs 6--21 times less CPU time than a leading local MP2 method at the same accuracy.

The remaining computational cost, and the $O(N^3)$ memory, belong to the input tensor preparation: the DF transform and the Laplace dressing, both dense $O(N^4)$ matrix multiplications, which produce the dressed tensors $\tilde{\boldsymbol{R}}(\beta)$ required by the present algorithm.
These tensors are local in the localized basis, so their preparation and storage can be reduced by utilizing locality, as in local correlation methods.
Since local approximations reduce the cost of an underlying algorithm by restricting it to local domains, their prefactor is set by the scaling of that algorithm within a domain. Applying them to the $O(N^4)$ input tensor preparation, rather than to the $O(N^5)$ evaluation of conventional MP2, can therefore be expected to give a considerably smaller prefactor.
One can further treat $\boldsymbol{U}^{\mathrm{o}}(\beta)$ and $\boldsymbol{U}^{\mathrm{v}}(\beta)$ as additional input tensors in an enlarged STC contraction, which can still be evaluated by our general loop-breaking strategy. The required input then reduces to the single undressed tensor $\boldsymbol{R}$, which is more local since it does not depend on the Fock matrix, and the dressing of $M$ tensors is avoided.
With such utilization of locality, STC-MP2 can in principle achieve a further reduced scaling, and eventually linear scaling, while retaining the small prefactor and the early crossover with DF-MP2 shown here. We leave this to future work.

The techniques in this work can be applied broadly beyond the evaluation of the MP2 energy, since MP2-like correlation terms enter regularized MP2 methods, double-hybrid density functionals, and higher-level local correlation approaches such as LNO-CCSD(T).\cite{lee2018,shee2021,rettig2022,carterfenk2023,grimme2006,martin2020,nagy2018}
Together with the previous application of STC to CCSD(T), the present results show that stochastic tensor contraction can serve as a general approach to reduce the cost of many-body methods in quantum chemistry.\cite{sun2026stc}

\section{Acknowledgments}
This work was supported by the U.S. Department of Energy, Office of Science, Basic Energy Sciences, under Award No.~DE-SC0018140. Large language models (Claude 5.0 and 5.5, Anthropic) were used to assist with coding, data analysis, and editing of the manuscript; all content was carefully checked by the authors, who take full responsibility for it.

\section*{Data Availability}
The code, numerical results, and plotting scripts that support the findings of this study are publicly available at \url{https://github.com/SUSYUSTC/stc_mp2_exchange}.

\end{document}